\documentclass[aps, amsmath, floats, floatfix, onecolumn, nofootinbib]{revtex4}
\usepackage{graphicx}
\usepackage{color}

\newcommand{\be}{\begin{eqnarray}}
\newcommand{\ee}{\end{eqnarray}}
\newcommand{\rar}{\rightarrow}

\begin{document}

\title{Rotating regular black holes}

\author{Cosimo Bambi}
\email{bambi@fudan.edu.cn}

\author{Leonardo Modesto}
\email{lmodesto@fudan.edu.cn}

\affiliation{Center for Field Theory and Particle Physics \& Department of Physics, Fudan University, 200433 Shanghai, China}

\date{\today}

\begin{abstract}
The formation of spacetime singularities is a quite common phenomenon in 
General Relativity and it is regulated by specific theorems. It is widely believed that 
spacetime singularities do not exist in Nature, but that they represent a limitation of 
the classical theory. While we do not yet have any solid theory of quantum gravity, 
toy models of black hole solutions without singularities have been proposed. 
So far, there are only non-rotating regular black holes in the literature. These metrics 
can be hardly tested by astrophysical observations, as the black hole spin plays a 
fundamental role in any astrophysical process. In this letter, we apply the 
Newman-Janis algorithm to the Hayward and to the Bardeen black hole metrics. 
In both cases, we obtain a family of rotating solutions. Every solution corresponds 
to a different matter configuration. Each family has one solution with special 
properties, which can be written in Kerr-like form in Boyer-Lindquist coordinates. 
These special solutions are of Petrov type D, they are singularity free, but they 
violate the weak energy condition for a non-vanishing spin and their curvature 
invariants have different values at $r=0$ depending on the way one approaches 
the origin. We propose a natural prescription to have rotating solutions with a 
minimal violation of the weak energy condition and without the questionable 
property of the curvature invariants at the origin. 
\end{abstract}

\maketitle


\section{Introduction}

Under the main assumptions of the validity of the strong energy condition and 
of the existence of global hyperbolicity, in General Relativity collapsing matter
forms spacetime singularities~\cite{sing}. At a singularity, predictability is lost and standard 
physics breaks down. In analogy with the appearance of divergent quantities 
in other classical theories, it is widely believed that even spacetime singularities
are a symptom of the limitations of General Relativity and that they must be 
solved in a theory of quantum gravity. While quantum gravity effects are traditionally
thought to show up at the Planck scale, $L_{\rm Pl} \sim 10^{-33}$~cm, making
experimental and observational tests likely impossible, more recent studies
have put forward a different idea~\cite{mathur,gia}: $L_{\rm Pl}$ would be the 
quantum gravity scale for a system of a few particles, while the quantum gravity
scale for systems with many constituents would be its gravitational radius. 
In these frameworks, even astrophysical black holes (BHs) of tens or millions
Solar masses may be intrinsically quantum objects, macroscopically different
from the Kerr BHs predicted in General Relativity.

While we do not yet have any mature and reliable candidate for a quantum theory 
of gravity, more phenomenological approaches have tried to somehow solve these 
singularities and study possible implications. In this context, an important line 
of research is represented by the work on the so-called regular BH 
solutions~\cite{bar,ag,hay,rbh}. These spacetimes have an event horizon and no
pathological features like singularities or regions with closed timelike curves. Of 
course, their metric is not a solution of Einstein's vacuum equations, but they 
can be introduced either with some exotic field, usually some form of non-linear 
electrodynamics, or modifications to gravity. They can avoid the singularity theorems 
because they meet the weak energy condition, but not the strong one.

The purpose of the present letter is to construct rotating regular BH solutions. This 
is a necessary step to test these metrics with astrophysical observations~\cite{rev,bb,b}. 
The spin enters as the current-dipole moment of the gravitational field of a compact 
object and it is thus the leading order correction to the mass-monopole term. It is not
possible to constrain deviations from classical predictions without an independent
estimate of the spin. However, exact rotating BH solutions different from the classical
Kerr-Newman metric are very hard to find. In most cases, including all the regular
BH metrics currently available in the literature, we know only the non-rotating
solution. In a few cases, we have an approximated solution valid in the slow-rotation
limit~\cite{slow}, which is also not very useful for tests. A rotating solution in the
Einstein-Gauss-Bonnet-dilaton gravity has been recently found numerically in 
Ref.~\cite{degb}, while proposals for some rotating quantum BHs have been 
suggested in~\cite{loop,belt}.

\section{Newman-Janis algorithm}

Our strategy is to use the Newman-Janis transformation~\cite{nj} (for more details, 
see Ref.~\cite{ds}). Roughly speaking, the algorithm starts with a non-rotating 
spacetime and, at the end of the procedure, the spacetime has an asymptotic 
notion of angular momentum. The starting point is a spherically symmetric spacetime
\be\label{eq-le}
ds^2 = f(r) dt^2 - \frac{dr^2}{f(r)} - h(r) \left(d\theta^2 + \sin^2\theta d\phi^2 \right) \, .
\label{GSS}
\ee
The {\it first step} of the algorithm is a transformation to get null coordinates 
$\left\{u,r,\theta, \phi\right\}$, 
where 
\be
du &=& dt - dr/f(r) \, .
\ee
The {\it second step} is to find a null tetrad $Z^\mu_\alpha = (l^\mu,n^\mu,m^\mu,\bar{m}^\mu)$ 
for the inverse matrix in null coordinates
\be\label{tetra}
g^{\mu \nu } = l^{\mu} n^{\nu} +  l^{\nu} n^{\mu} -  m^{\mu} {\bar m}^{\nu} 
-  m^{\nu} {\bar m}^{\mu} \, ,
\ee
where the tetrad vectors satisfy the relations
\be
l_{\mu} l^{\mu} = m_{\mu} m^{\mu} = n_{\mu} n^{\mu} = l_{\mu} m^{\mu} 
=n_{\mu} m^{\mu} =0 \, , \quad
l_{\mu} n^{\mu} = - m_{\mu} {\bar m}^{\mu} =1 \, ,
\ee
and ${\bar x}$ is the complex conjugate of the general quantity $x$. One finds
\be
l^{\mu } = \delta_r^{\mu} \, , \quad
n^{\mu} = \delta^{\mu}_u - \frac{f(r)}{2} \delta^{\mu}_r \, , \quad
m^{\mu} = \frac{1}{\sqrt{2 h(r)}} \left( \delta_{\theta}^{\mu} + \frac{i}{\sin\theta} 
\delta_{\phi}^{\mu} \right) \, .
\ee
The {\it third step} of the procedure is the combination of two operations. A complex 
transformation in the $r-u$ plane as follows
\begin{eqnarray}\label{NJ} 
r \rightarrow r' = r + i \, a \cos \theta \, , \quad
u \rightarrow& u' = u  - i \, a \cos \theta \, ,
\end{eqnarray}
together with a complexification of the functions $f(r)$ and $h(r)$ of the metric. 
The new tetrad vectors are
\be\label{tetra2}
l'^{\mu } = \delta_r^{\mu} \, , \quad
n'^{\mu} =\delta^{\mu}_u - \frac{\tilde{f}(r')}{2} \delta^{\mu}_r \, , \quad
m'^{\mu} = \frac{1}{\sqrt{ 2 \tilde h(r')}}  
\left(i a \sin\theta (\delta^{\mu}_u - \delta^{\mu}_r) + \delta_{\theta}^{\mu} 
+ \frac{ i}{\sin \theta} \delta_{\phi}^{\mu} \right) \, ,
\ee
where $\tilde{f}(r')$ and $\tilde{h}(r')$ are real functions on the complex domain. 
This step of the procedure is in principle completely arbitrary. In fact, in the original 
paper, Newman and Janis could not give a true explanation of the procedure if 
not that it works for the Kerr metric with a particular choice of the complexifications. 
The situation improved with Drake and Szekeres in~\cite{ds}, where the authors
proved that the only Petrov~D spacetime generated by the Newman-Janis algorithm 
with a vanishing Ricci scalar is the Kerr-Newman solution. Using the new tetrad in
Eq.~(\ref{tetra}), we find the new inverse metric and then the metric. The non-vanishing
coefficients of $g_{\mu\nu}$ are
\be
g_{uu} &=& \tilde{f}(r,\theta) \, , \quad g_{ur} = g_{ru} = 1 \, , \quad 
g_{u\phi} = g_{\phi u} = a \sin^2\theta \left(1 - \tilde{f}(r,\theta) \right) \, , \nonumber\\
g_{r\phi} &=& g_{\phi r} = a \sin^2\theta \, , \quad
g_{\theta\theta} = - \tilde{h}(r,\theta) \, , \quad
g_{\phi\phi} = - \sin^2\theta \left[\tilde{h}(r,\theta) + 
a^2 \sin^2\theta \left(2 - \tilde{f}(r,\theta) \right) \right] \, .
\ee
The {\it fourth and last step} of the algorithm is a change of coordinates. In some cases,
we can write the metric in the Boyer-Lindquist form, in which the only non-vanishing 
off-diagonal term is $g_{t\phi}$. This requires a coordinate transformation of the form
\be\label{eq-tr}
du = dt' + F(r) dr \, , \quad d\phi = d\phi' + G(r) dr \, ,
\ee 
where
\be\label{eq-FG}
F(r) = \frac{\tilde{h}(r,\theta) + a^2 \sin^2\theta}{\tilde{f}(r,\theta) 
\tilde{h}(r,\theta) + a^2 \sin^2\theta} \, , \quad
G(r) = \frac{a}{\tilde{f}(r,\theta) \tilde{h}(r,\theta) + a^2 \sin^2\theta} \, .
\ee
This transformation is possible only when $F$ and $G$ depend on the coordinate
$r$ only. In general, however, the expressions on the right hand sides of~(\ref{eq-FG})
depend also on $\theta$, and we cannot perform a global transformation of the
form~(\ref{eq-tr}). If the transformation~(\ref{eq-tr}) is allowed and we go to 
Boyer-Lindquist coordinates, the non-vanishing metric coefficients of the rotating 
BH metric are:
\be\label{eq-rot}
&&g_{tt} = \, \tilde{f}(r,\theta) \, , \quad
g_{t\phi} = g_{\phi t} = a \sin^2\theta \left(1 - \tilde{f}(r,\theta)\right) \, , \quad 
g_{rr} = - \frac{\tilde{h}(r,\theta)}{\tilde{h}(r,\theta) \tilde{f}(r,\theta) 
+ a^2 \sin^2\theta} \, , \nonumber\\
&&g_{\theta\theta} = - \tilde{h}(r,\theta) \, , \quad
g_{\phi\phi} = - \sin^2\theta \left[ \tilde{h}(r,\theta) + a^2 
\sin^2\theta \left(2 - \tilde{f}(r,\theta)\right) \right] \, .
\ee

In the case of the Schwarzschild solution, we have $f(r) = 1 - 2M/r$ and $h(r) = r^2$. 
In the Newman-Janis algorithm, we have to choose a complexification of the $1/r$ 
and of the $r^2$ term. In general, this prescription is not unique. However, since 
we know what the Kerr solution is, we know that if we take the following 
complexification
\be\label{schwarz}
\frac{1}{r} \rightarrow \frac{1}{2} \left(\frac{1}{r'}+\frac{1}{\bar{r}'} \right) \, , 
\quad r^2 \rightarrow r' \bar{r}' \, , 
\ee
then this trick works well. The functions $f(r)$ and $h(r)$ become
\begin{eqnarray}\label{schwarzschild}
f(r) \rar  \tilde{f}(r,\theta) = 1 - \frac{2Mr}{\Sigma} \, , \quad
h(r) \rar \tilde{h}(r,\theta) = \Sigma \, ,
\end{eqnarray}
where $\Sigma = r^2 + a^2\cos^2\theta$. In this case, the functions $F$ and $G$ in
Eq.~(\ref{eq-FG}) depend on $r$ only and we find the Kerr solution in
Boyer-Lindquist coordinates
\be\label{kerr}
ds^2 = \left(1 - \frac{2 M r}{\Sigma}\right) dt^2 
+ \frac{4 a M r \sin^2\theta}{\Sigma} dt d\phi
- \frac{\Sigma}{\Delta} dr^2 
- \Sigma d\theta^2 - \sin^2 \theta \left(r^2 + a^2 
+ \frac{2 a^2 M r \sin^2\theta}{\Sigma} \right) d\phi^2 \, ,
\ee
where $\Delta = r^2 - 2 M r + a^2$.

\section{Hayward black hole}

As first example of regular black hole, we consider the Hayward metric, whose
analytic expression is quite simple~\cite{hay}. The line element is given by 
Eq.~(\ref{eq-le}), with the following $f(r)$  
\be\label{eq-hay}
f(r) = 1 - \frac{2 m}{r} \, , \quad m = m(r) = M \frac{r^3}{r^3 + g^3} \, ,
\ee 
where $M$ is the BH mass and $g$ is some real positive constant measuring 
the deviations from the classical Kerr metric. Let us note that $m(r)$ may be interpreted 
as the mass inside the sphere of radius $r$ and approaches $M$ as $r$ goes to infinity. 
This spacetime is everywhere regular, as can be verified by its curvature invariants:
\be
R = \frac{12 M g^3 \left(r^3 - 2 g^3\right)}{\left(r^3 + g^3\right)^3} \, , &&\quad
\lim_{r\rar0} R = - \frac{24 M}{g^3} \, , \\
R_{\mu\nu} R^{\mu\nu} = \frac{72 M^2 g^6 \left(5 r^6 - 3r^3g^3 
+ 2g^6\right)}{\left(r^3 + g^3 \right)^6} \, , &&\quad 
\lim_{r\rar0} R_{\mu\nu} R^{\mu\nu} = \frac{144 M^2}{g^6} \, , \\
R_{\mu\nu\rho\sigma} R^{\mu\nu\rho\sigma} = 
\frac{48 M^2 \left(r^{12} - 4 r^9g^3 + 18r^6g^6 -2r^3g^9 
+ 2g^{12}\right)}{\left(r^3 + g^3 \right)^6} \, , &&\quad
\lim_{r\rar0} R_{\mu\nu\rho\sigma} R^{\mu\nu\rho\sigma} = 
\frac{96 M^2}{g^6} \, .
\ee
The weak energy condition is also satisfied.

When we apply the Newman-Janis algorithm to get a rotating solution, the key-point 
is the complexification of $f(r)$, as the one for $h(r)$ must be the same of 
Eq.~(\ref{schwarzschild}). With the only requirement to recover the Kerr metric 
for $g=0$, all the possible complexifications have the form
\be
\tilde{f} (r,\theta) = 1 - \frac{2 \tilde{m} r}{\Sigma} \, ,
\ee
where, in general, $\tilde{m} = \tilde{m}_{\alpha,\beta}(r,\theta)$ is a function of
both $r$ and $\theta$ and the complexification is characterized by the two real
numbers $\alpha$ and $\beta$:
\be
\tilde{m}_{\alpha,\beta}(r,\theta) 
= M \frac{r^{3+\alpha} \Sigma^{-\alpha/2}}{r^{3+\alpha}
\Sigma^{-\alpha/2} + g^3 r^{\beta} \Sigma^{-\beta/2}} \, .
\ee

At this point, we can distinguish two classes of solutions. The first class has only 
the case $\alpha = \beta = 0$ (complexification of type-I): we complexify the 
$1/r$ term as in Schwarzschild, without altering the mass term $m(r)$. With 
this choice
\be\label{eq-c1}
\tilde{f}_I (r,\theta) = 1 - \frac{2 m r}{\Sigma} \, ,
\ee
$F$ and $G$ depend on the coordinate $r$ only, and the final result is the
line element~(\ref{kerr}) with $m(r)$ in Eq.~(\ref{eq-hay}) replacing $M$.
The spacetime is of Petrov type~D, as we can verify by the presence of the
Carter constant for the motion of a free particle. The new solution is also
everywhere regular for $g\neq0$, as can be seen from the expression of
its curvature scalar, Ricci square, and Kretschmann invariant. We do not 
report the analytic forms here, but they can be quickly obtained with Mathematica 
and any good package for tensor calculations. At the origin, they reduce to
\be
\lim_{\theta\rar {\rm any}} \left( \lim_{r\rar0} R \right) = 
\lim_{r\rar0} \left( \lim_{\theta\rar\theta\neq\pi/2} R \right) = 0 \, , && \quad
\lim_{r\rar0} \left( \lim_{\theta\rar\pi/2} R \right) =  - \frac{24 M}{g^3} \, , \\
\lim_{\theta\rar {\rm any}} \left( \lim_{r\rar0} 
R_{\mu\nu} R^{\mu\nu} \right) = \lim_{r\rar0} \left( \lim_{\theta\rar\theta\neq\pi/2} 
R_{\mu\nu} R^{\mu\nu} \right) = 0 \, , && \quad
\lim_{r\rar0} \left( \lim_{\theta\rar\pi/2} R_{\mu\nu} R^{\mu\nu} \right) = 
\frac{144 M^2}{g^6} \, , \\
\lim_{\theta\rar {\rm any}} \left( \lim_{r\rar0} 
R_{\mu\nu\rho\sigma} R^{\mu\nu\rho\sigma} \right) = 
\lim_{r\rar0} \left( \lim_{\theta\rar\theta\neq\pi/2} R_{\mu\nu\rho\sigma} 
R^{\mu\nu\rho\sigma} \right) = 0 \, , && \quad
\lim_{r\rar0} \left( \lim_{\theta\rar\pi/2} R_{\mu\nu\rho\sigma} 
R^{\mu\nu\rho\sigma} \right) = \frac{96 M^2}{g^6} \, .
\ee
The fact that these curvature invariants assume two different values for $r=0$, 
depending on the way one approaches the origin, is a
signature of the ``de Sitter belt'', absent in the non-rotating metric. It has been 
already found in some non-commutative geometry inspired BHs~\cite{belt}.

\begin{figure}
\begin{center}
\includegraphics[type=pdf,ext=.pdf,read=.pdf,width=7cm]{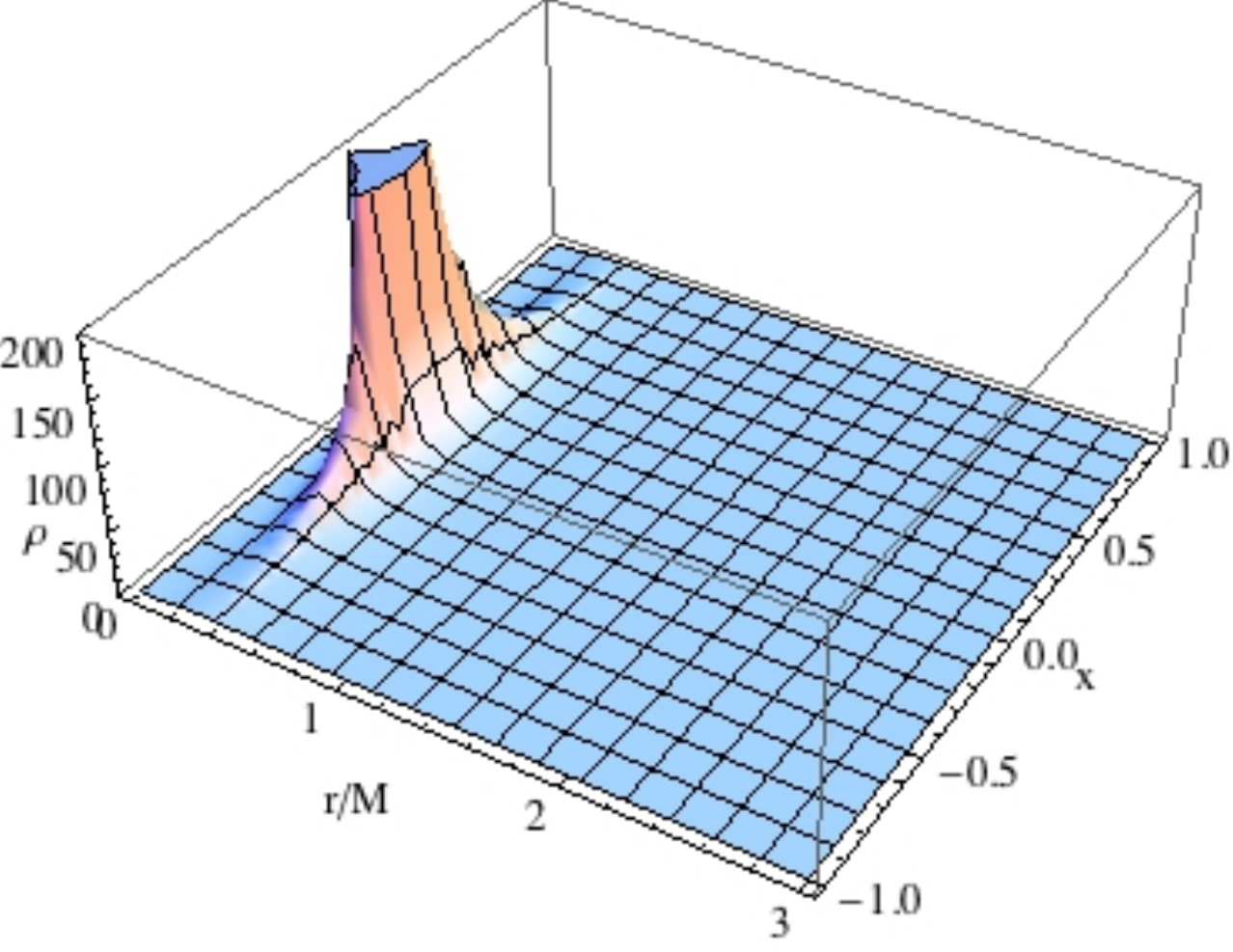} \hspace{0.8cm}
\includegraphics[type=pdf,ext=.pdf,read=.pdf,width=7cm]{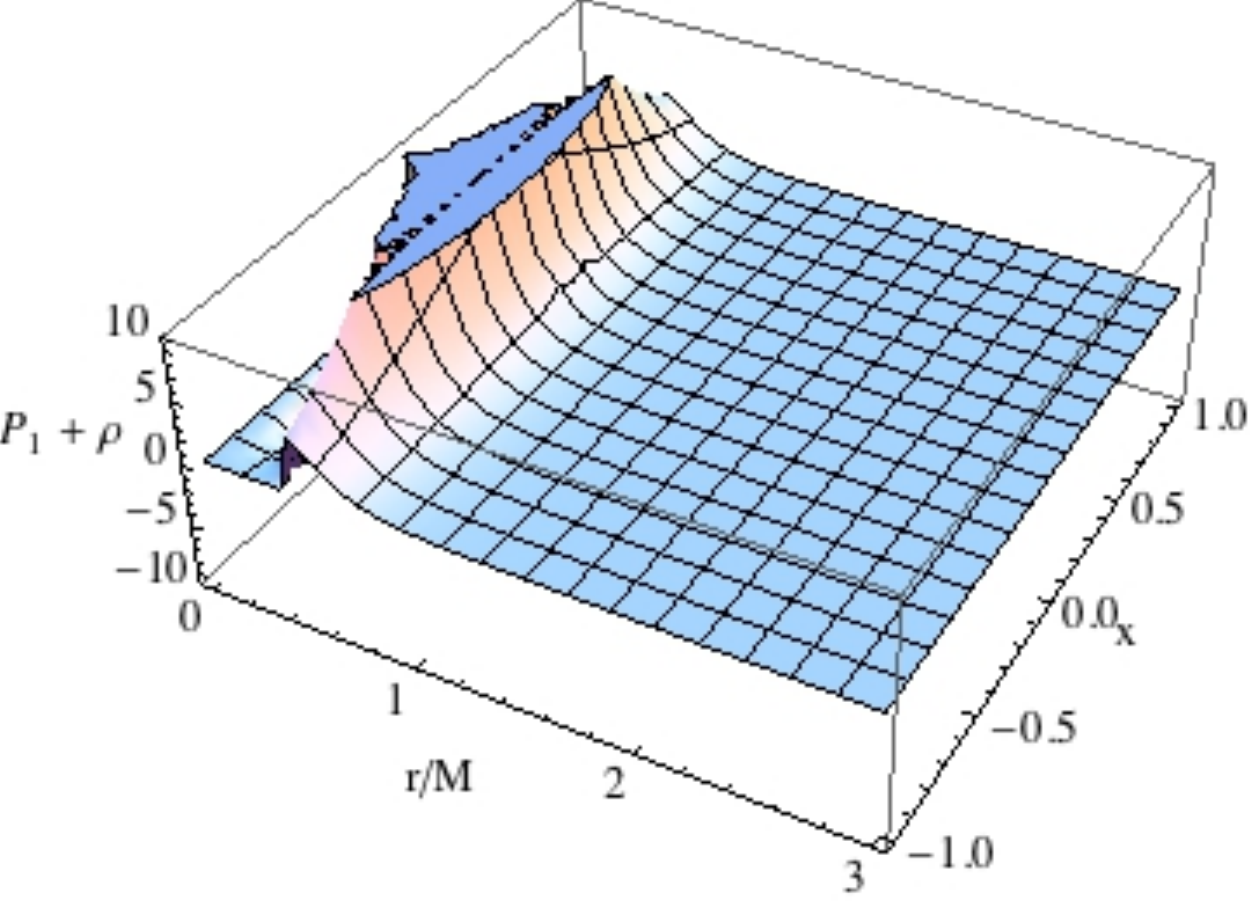} 
\includegraphics[type=pdf,ext=.pdf,read=.pdf,width=7cm]{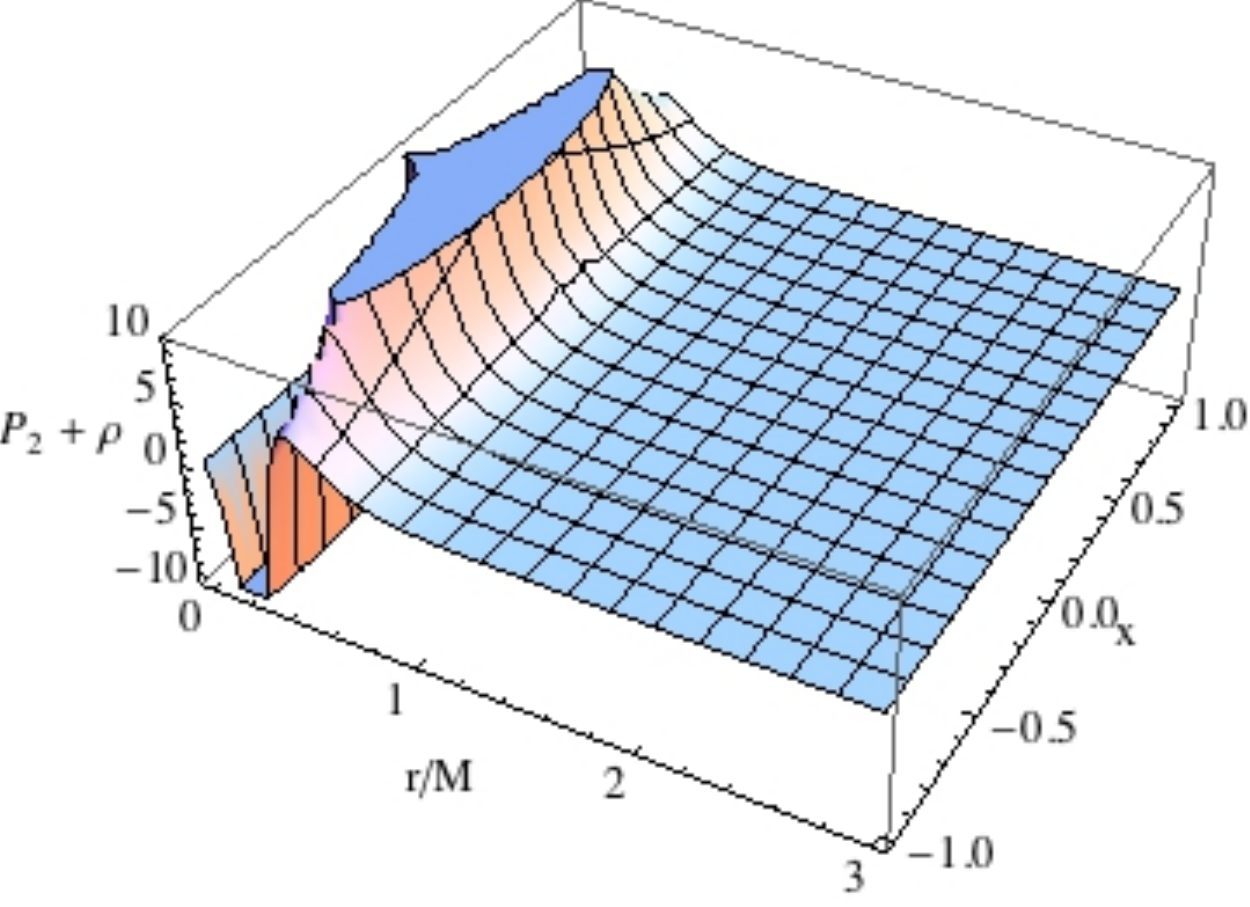} \hspace{0.8cm}
\includegraphics[type=pdf,ext=.pdf,read=.pdf,width=7cm]{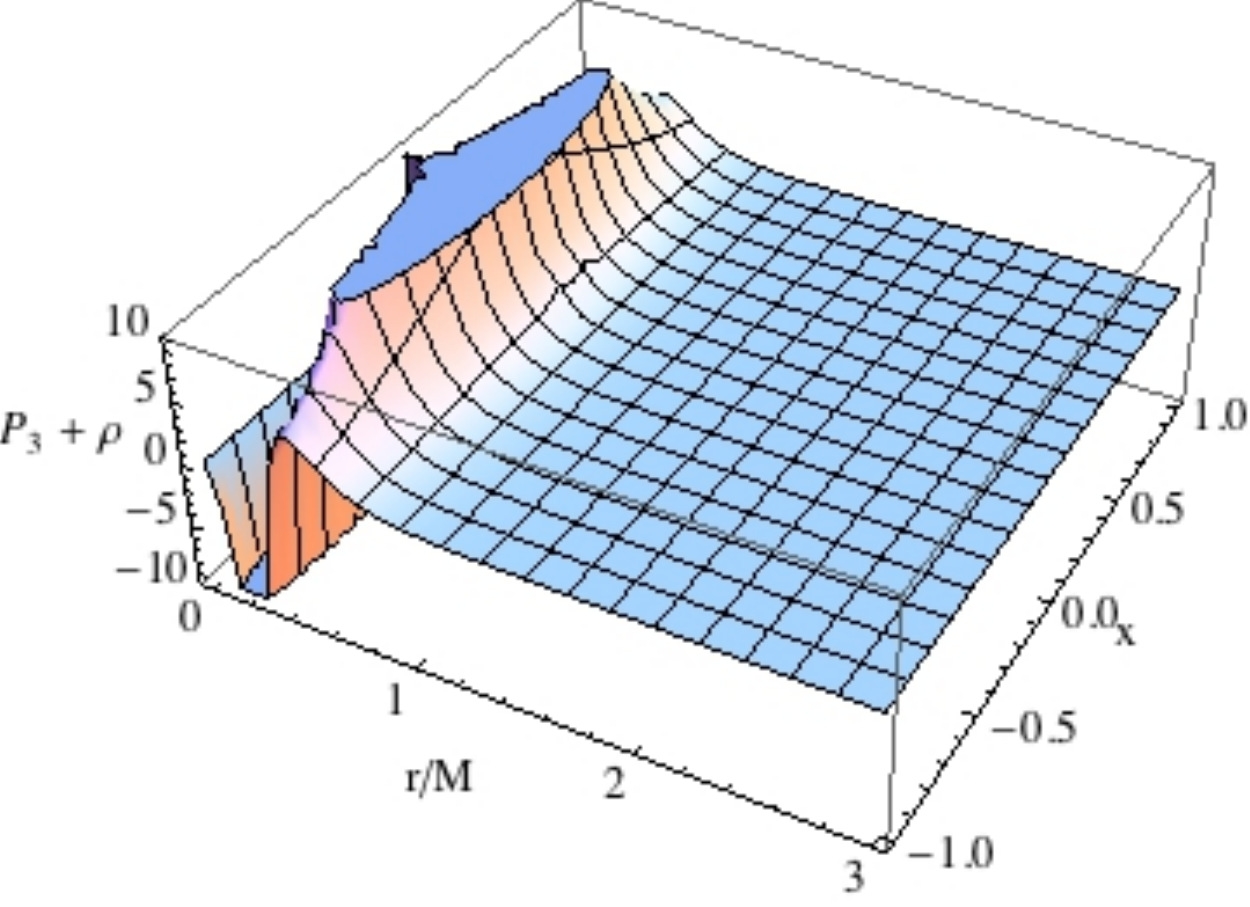}
 \end{center}
\caption{Plot of $\rho$ (top left panel), $\rho+P_1$ (top right panel),
$\rho+P_2$ (bottom left panel), and $\rho+P_3$ (bottom right panel) for 
a rotating Hayward BH with $a/M = 0.6$ and $g = 0.3$. Here $x = \cos\theta$.}
\label{f1}
\end{figure}

The weak energy condition, satisfied in the non-rotating case, is violated for 
$a\neq0$. To check it, we can choose an orthonormal basis in which the 
stress-energy tensor is diagonal, $T^{(a)(b)} = {\rm diag}(\rho,P_1,P_2,P_3)$.
The weak energy condition requires $\rho \ge 0$ and $\rho + P_i \ge 0$ 
$(i=1,2,3)$~\cite{sing}. The one forms of the dual basis of the orthonormal 
tetrad of the standard locally non-rotating frame are~\cite{teu}
\be
{\bf e}^{(0)} = \left| g_{tt} - \frac{g_{t\phi}^2}{g_{\phi\phi}}\right|^{1/2} {\bf dt} \, , \,\,
{\bf e}^{(1)} = \left| g_{rr} \right|^{1/2} {\bf dr} \, , \,\,
{\bf e}^{(2)} = \left( - g_{\theta\theta} \right)^{1/2} {\bf d\theta} \, , \,\,
{\bf e}^{(3)} = - \frac{g_{t\phi}}{\left( - g_{\phi\phi} \right)^{1/2}} {\bf dt} + 
\left( - g_{\phi\phi} \right)^{1/2} {\bf d\phi}\, .
\ee
In this frame, the Einstein tensor has an off-diagonal element $G^{(0)(3)}$, so
we need a transformation $({\bf e}^{(0)},{\bf e}^{(3)}) \rar ({\bf e}'^{(0)},{\bf e}'^{(3)})$ 
to get a diagonal tensor. At this point, the density $\rho$ is the eigenvalue of the
timelike eigenvector (which is ${\bf e}'^{(0)}$ outside the BH, ${\bf e}^{(1)}$
between the outer and inner horizon, and ${\bf e}'^{(3)}$ inside the inner horizon),
while $P_1$, $P_2$, and $P_3$, corresponding to the principal pressures
in the three spacelike directions, are the three eigenvalues of the spacelike
eigenvectors. Fig.~\ref{f1} shows the violation of the weak energy condition
for a BH with $a/M=0.6$ and $g=0.3$.

The second class of transformations (complexification of type-II) includes all 
the other options. Now $\tilde{m}$ is a function of both $r$ and $\theta$, except 
in the Kerr limit $g=0$. The common feature of all these solutions is that there 
is no global transformation~(\ref{eq-tr}) to write the new metric in the Kerr form 
in Boyer-Lindquist coordinates. That can happen because we are not in the 
vacuum (assuming the validity of Einstein's equations and introducing 
some form of exotic matter like a non-linear electrodynamics field) and the 
stress-energy tensor is not the one of a Maxwell electromagnetic field, so the 
hypothesis $(ii)$ of Theorem~7.1.1 of Ref.~\cite{wald} is not satisfied.  
However, we can still check the absence of singularities and the validity of
the weak energy conditions, as they are both local properties. The
metric is now more complicated. We have just considered the case $\alpha = -2$ 
and $\beta = 0$. Like for the case $\alpha=\beta=0$, it turns out that the 
spacetime is everywhere regular for $g\neq0$, the limit of the curvature 
invariants assumes different values at $r=0$ depending on the way one 
approaches the origin, and the weak energy condition is not respected for
a non-vanishing spin $a$.

\section{Bardeen black hole}

The most famous regular BH solution is the Bardeen metric~\cite{bar}. In 
Schwarzschild coordinates, the line element is given by Eq.~(\ref{eq-le}) with
\be\label{eq-bar}
f(r) = 1 - \frac{2 m}{r} \, , \quad m = m(r) = 
M \left(\frac{r^2}{r^2 + g^2}\right)^{3/2} \, ,
\ee 
where $M$ is the mass of the BH and $g$ is its magnetic charge. The non-linear
electrodynamics field to obtain this metric from Einstein's equations was
found in Ref.~\cite{ag}. Even in this case, $m(r)$ may be interpreted as the 
mass inside the sphere of radius $r$. The spacetime is everywhere regular
(the analytic expression of the curvature invariants is reported in~\cite{ag})
and the weak energy condition is satisfied~\cite{ag}.

With the only requirement to recover the Kerr metric for $g=0$, all the possible
complexifications have the form
\be
\tilde{f} (r,\theta) = 1 - \frac{2 \tilde{m} r}{\Sigma} \, , \quad
\tilde{m}_{\alpha,\beta}(r,\theta) 
= M \left(\frac{r^{2+\alpha} \Sigma^{-\alpha/2}}{r^{2+\alpha}
\Sigma^{-\alpha/2} + g^2 \Sigma^{-\beta/2} r^{\beta}}\right)^{3/2} \, ,
\ee
which is quite similar to the case of the previous section. Once again, we have
two classes of solutions. The type-I solution has $\alpha = \beta = 0$ and corresponds
to the trivial complexification in which the mass term $m(r)$ is not modified. 
Such a solution can be written in the Kerr form in Boyer-Lindquist coordinates
with $m(r)$ in Eq.~(\ref{eq-bar}) replacing $M$, and it is of Petrov type~D.
All the other complexifications lead to solutions of type-II, in which we do not
recover the Boyer-Lindquist form of the metric. Both type-I and type-II BHs seem
to be everywhere singularity free (that is true for the type-I solution, we
have checked it is true even for some type-II, and we guess it is true for all
these metrics). The curvature invariants always assume different values for
$r=0$, depending on the way we approach the origin. For instance, the type-I
solution has
\be
\lim_{r\rar0} \left( \lim_{\theta\rar\pi/2} R \right) =  - \frac{24 M}{|g|^3} \, , \quad
\lim_{r\rar0} \left( \lim_{\theta\rar\pi/2} R_{\mu\nu} R^{\mu\nu} \right) = 
- \frac{144 M^2}{g^6} \, , \quad
\lim_{r\rar0} \left( \lim_{\theta\rar\pi/2} R_{\mu\nu\rho\sigma} 
R^{\mu\nu\rho\sigma} \right) = - \frac{96 M^2}{g^6} \, ,
\ee
while the limit is zero otherwise. The weak energy condition of all these solutions
is violated when $a \neq 0$.

\section{Revising the Newman-Janis algorithm}

On the base of the results presented in the previous sections, it seems like 
the Newman-Janis algorithm preserves the singularity free property of the
non-rotating solution, but not the weak energy condition. Moreover, a
peculiar feature concerning the limit $r\rar0$ of the curvature invariants 
shows up. These two properties are not very appealing and we may 
think about the way to avoid them.

As already pointed out in the previous sections, the quantity $m(r)$ in
Eqs.~(\ref{eq-hay}) and (\ref{eq-bar}) looks like the mass inside the sphere 
of radius $r$ and it reduces to $M$, the BH mass, at large radii. In the 
third step of the Newman-Janis algorithm, we introduce the quantity 
$a$ by hand, see Eq.~(\ref{NJ}), and we then identify such a parameter 
with the specific spin angular momentum of the BH, i.e. $a = J/M$.
In the standard case of Schwarzschild and Kerr BHs, in which the mass
is concentrated at the origin $r=0$, there are no ambiguities. However,
for all the regular BHs in the literature the mass of the object seems to 
be smeared over a larger volume, and it is thus questionable that the
specific spin is independent of $r$.

\begin{figure}
\begin{center}
\includegraphics[type=pdf,ext=.pdf,read=.pdf,width=7cm]{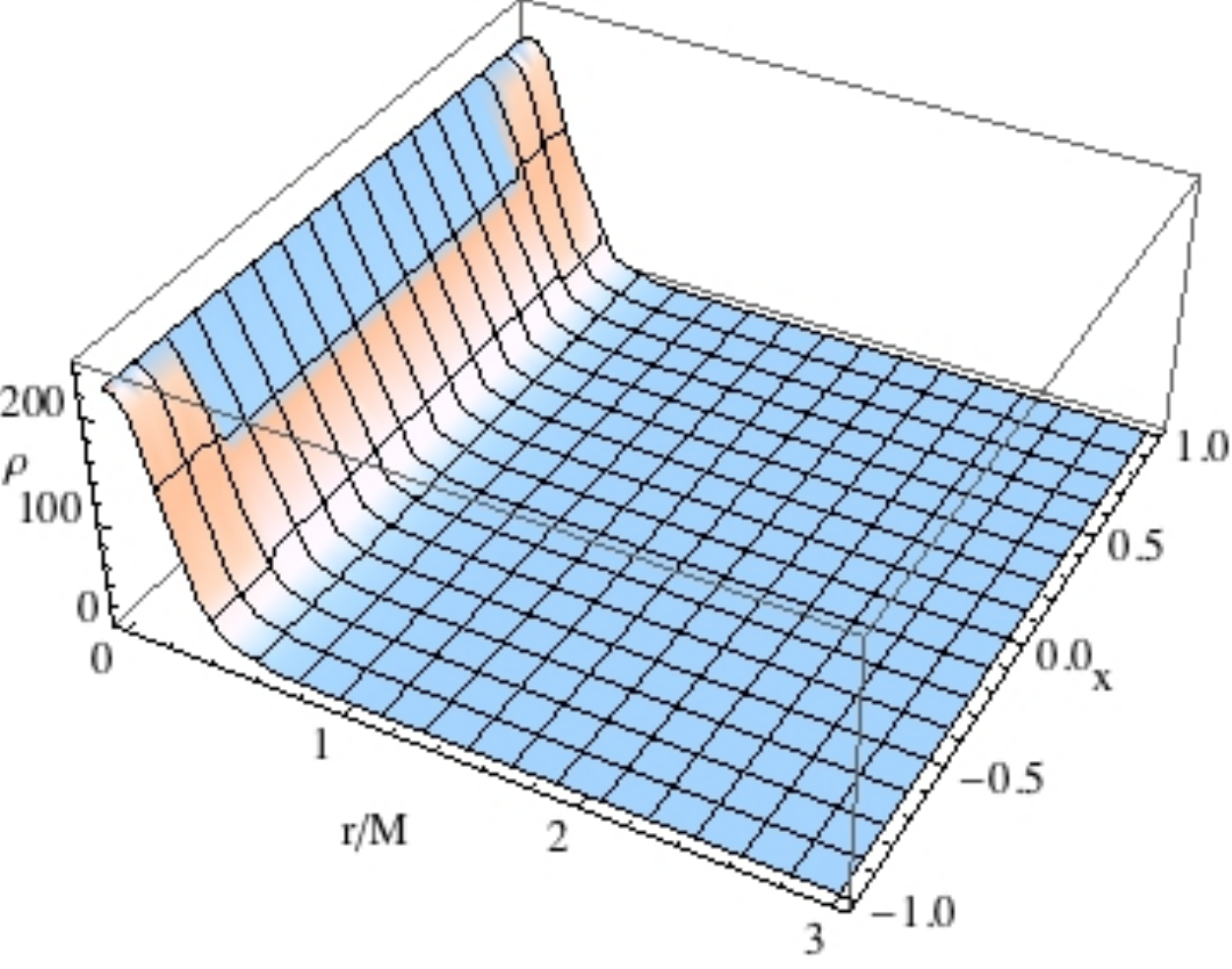} \hspace{0.8cm}
\includegraphics[type=pdf,ext=.pdf,read=.pdf,width=7cm]{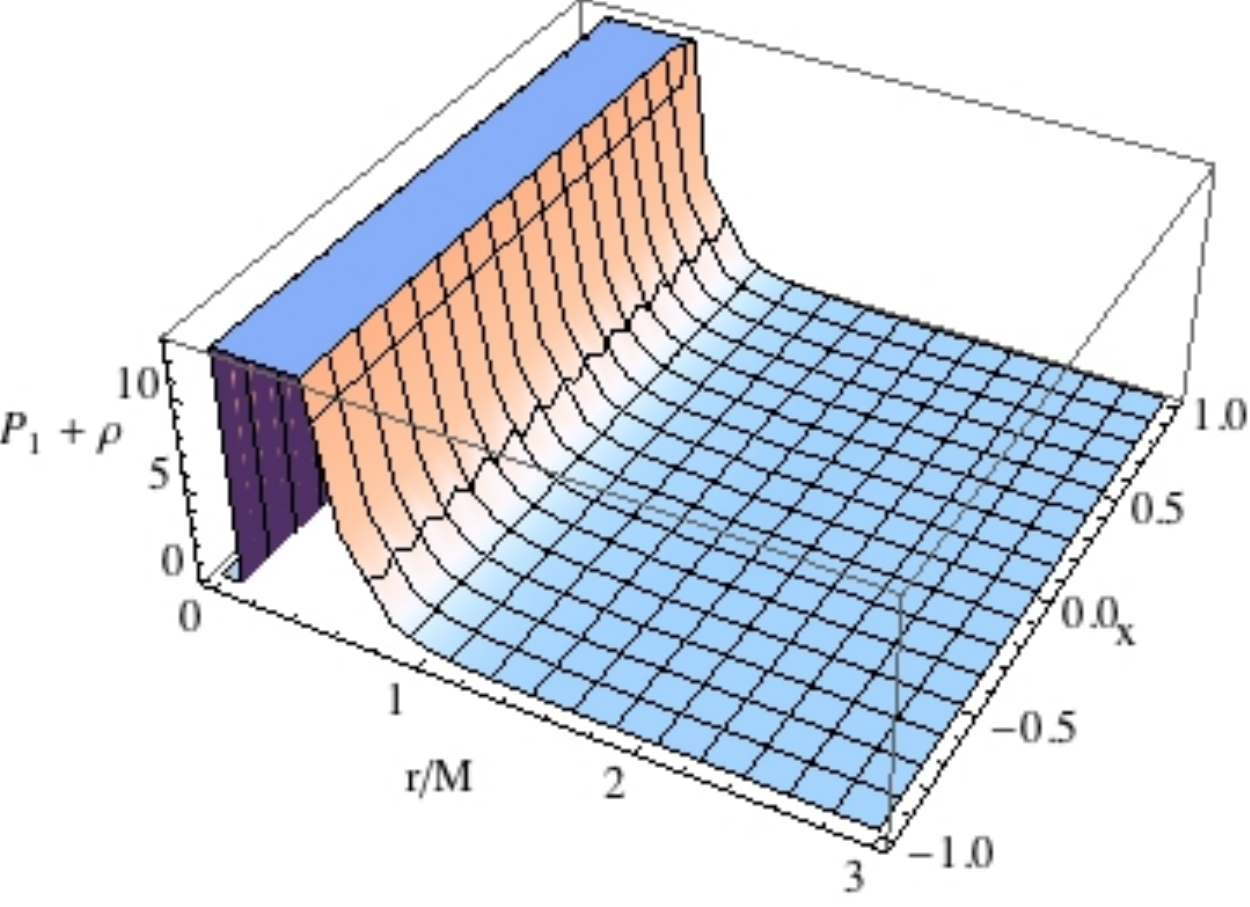}
\includegraphics[type=pdf,ext=.pdf,read=.pdf,width=7cm]{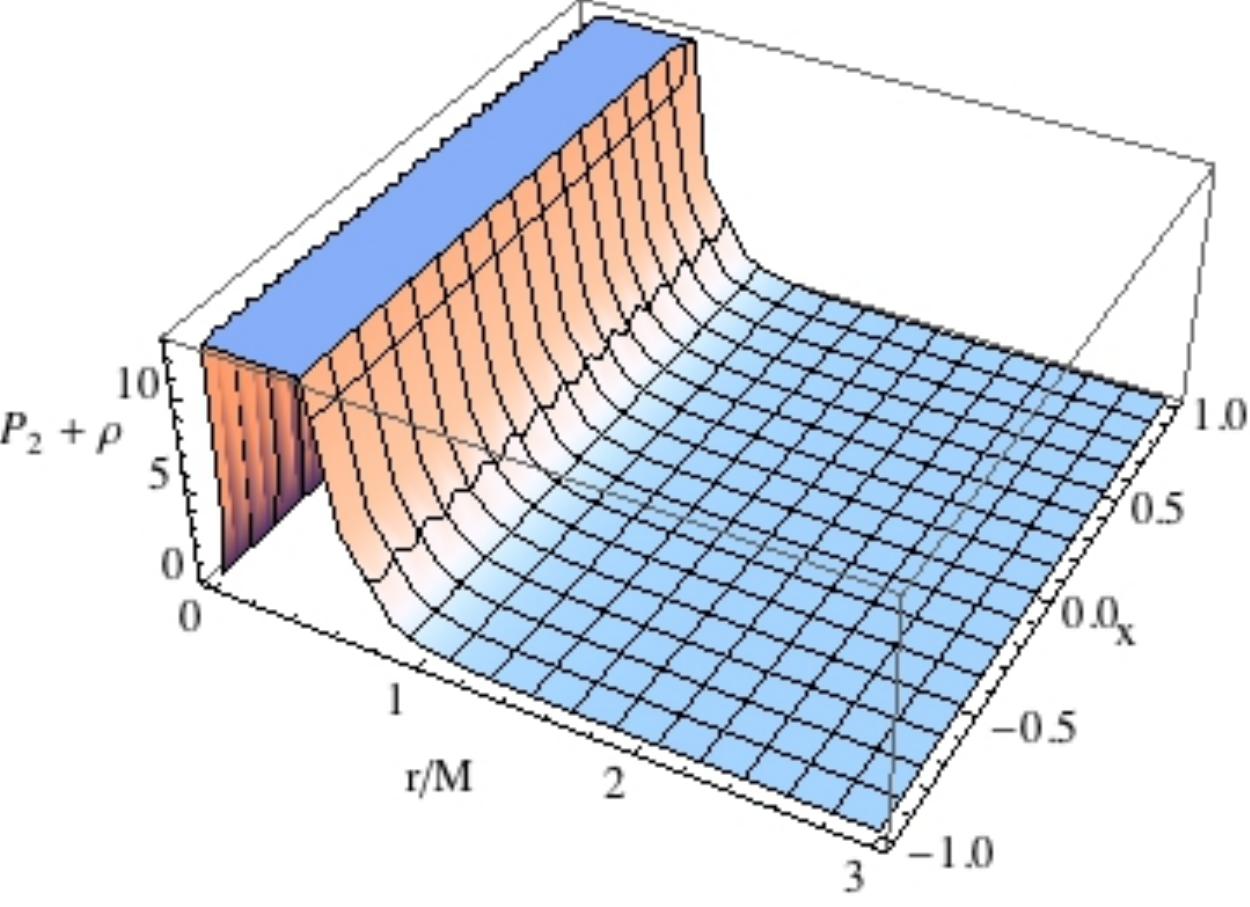} \hspace{0.8cm}
\includegraphics[type=pdf,ext=.pdf,read=.pdf,width=7cm]{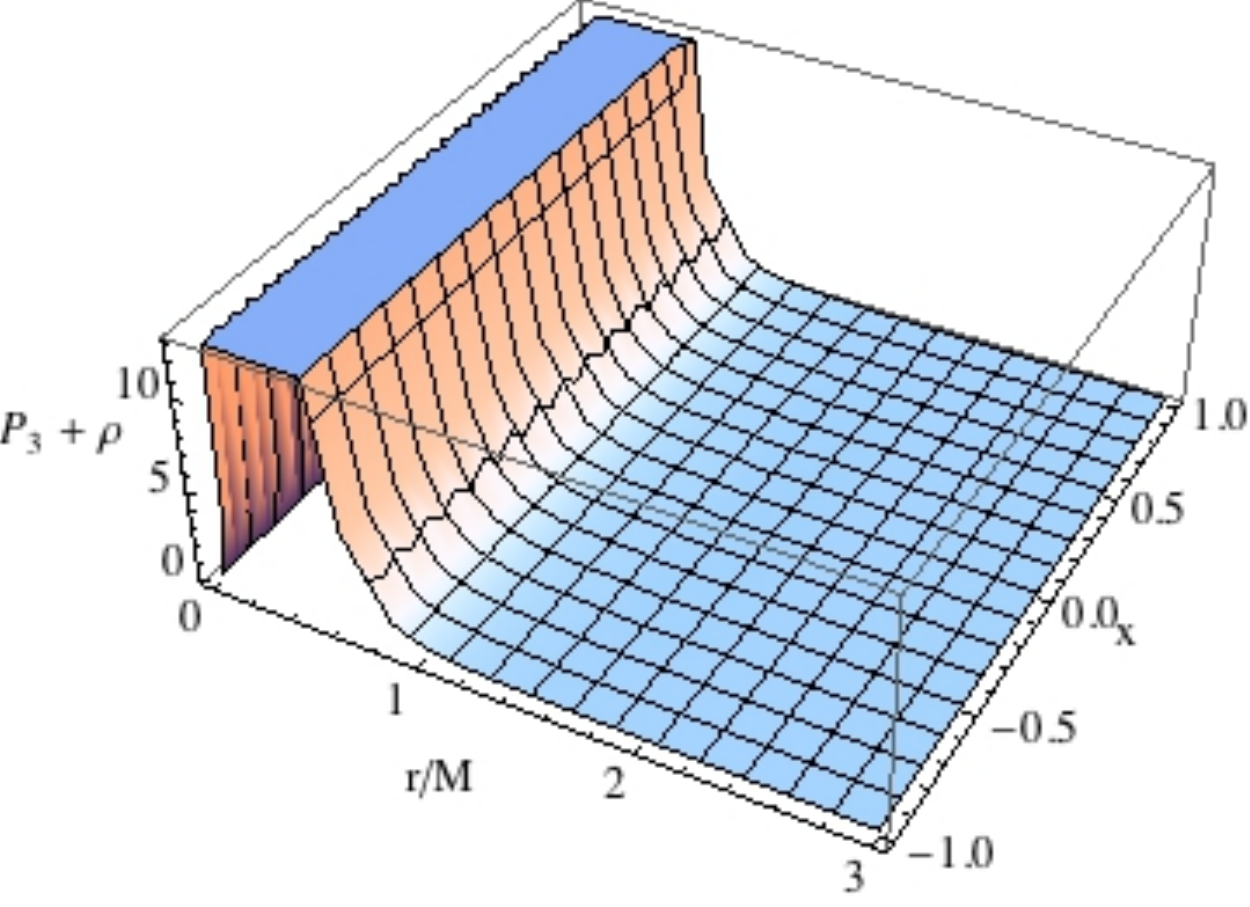}
 \end{center}
\caption{Plot of $\rho$ (top left panel), $\rho+P_1$ (top right panel),
$\rho+P_2$ (bottom left panel), and $\rho+P_3$ (bottom right panel) for 
a rotating Hayward BH with $a/M = 0.6$, $g = 0.3$, and $g' = 2.0$.
Here $x = \cos\theta$.}
\label{f2}
\end{figure}

\begin{figure}
\begin{center}
\includegraphics[type=pdf,ext=.pdf,read=.pdf,width=7cm]{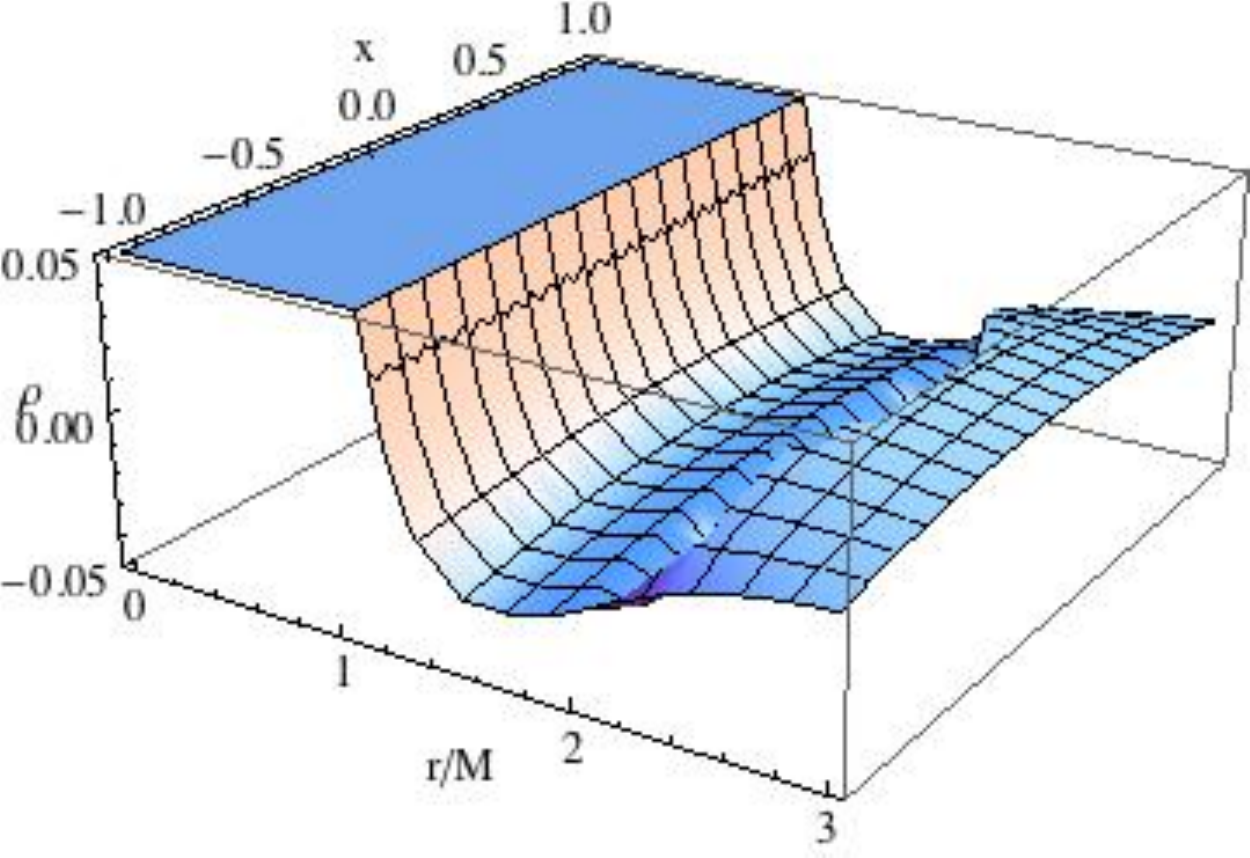} \hspace{0.8cm}
\includegraphics[type=pdf,ext=.pdf,read=.pdf,width=7cm]{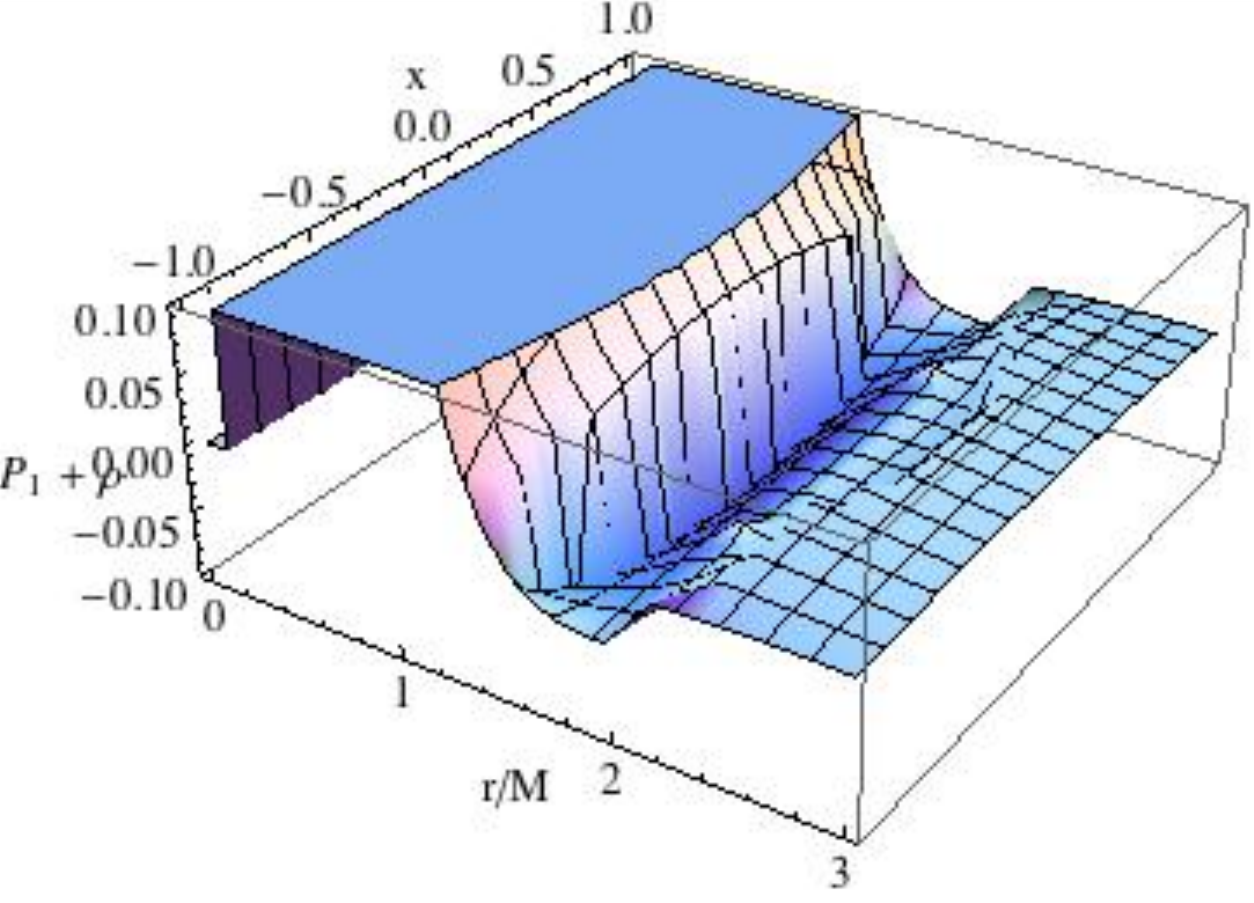}
\includegraphics[type=pdf,ext=.pdf,read=.pdf,width=7cm]{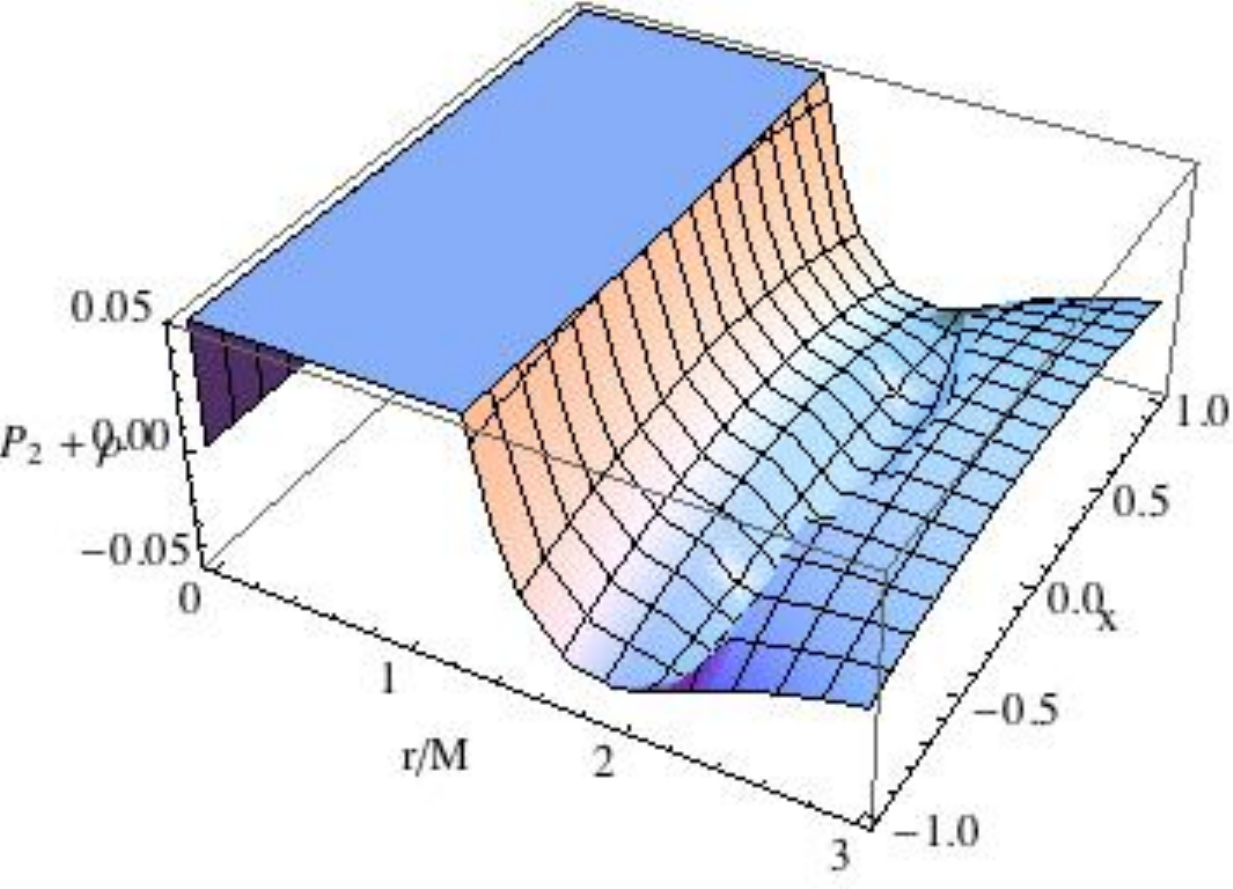} \hspace{0.8cm}
\includegraphics[type=pdf,ext=.pdf,read=.pdf,width=7cm]{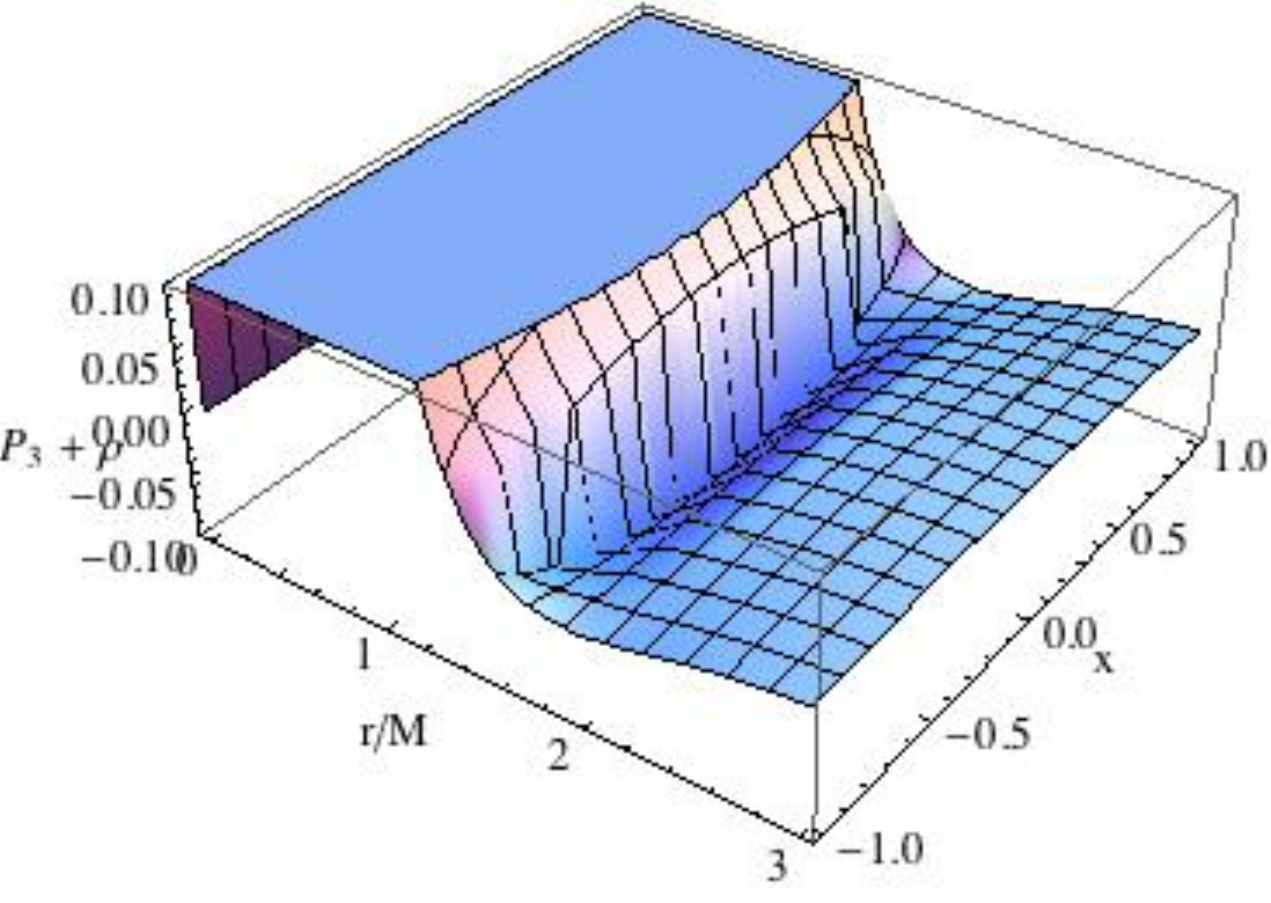}
 \end{center}
\caption{As in Fig.~\ref{f2}, but with a different range of the vertical axis to
show the small violation of the weak energy condition.}
\label{f2bis}
\end{figure}

We have tried to apply this idea to the type-I rotating Hayward BH
solution, as it is the one with the simplest analytical form. This passage is
clearly arbitrary, as we do not see any natural choice for $a(r)$ and
we do not know when the spin parameter should be promoted to the
rank of function. We have thus considered the simplest option,
in which the rotating BH solution looks like the Kerr metric
\be
ds^2 = \left(1 - \frac{2 m r}{\Sigma}\right) dt^2 
+ \frac{4 a' m r \sin^2\theta}{\Sigma} dt d\phi
- \frac{\Sigma}{\Delta} dr^2 
- \Sigma d\theta^2 - \sin^2 \theta \left(r^2 + a'^2 
+ \frac{2 a'^2 m r \sin^2\theta}{\Sigma} \right) d\phi^2 \, ,
\ee
with $m$ given in Eq.~(\ref{eq-hay}) and $a'$ given by
\be\label{eq-ap}
a' = a \frac{r^3}{r^3 + g'^3} \, ,
\ee
where $a=a'(r\rar\infty)=J/M$ is the specific spin angular momentum
at large radii and $g'$ is a new constant with a role similar to $g$. This
choice of $a'$ is just an example. The new spacetime has the curvature
invariants at the origin assuming the same values of the non-rotating
solution, independently of the way one approaches the point $r=0$.
The weak energy condition is not really satisfied, but the violation can be
very small, depending on the value of $g'$, as shown in Figs.~\ref{f2} and
\ref{f2bis} for a rotating Hayward BH with $a/M = 0.6$, $g = 0.3$, and $g' = 2.0$.

\section{Summary and Conclusions}

In this letter, we have applied the Newman-Janis algorithm to the regular Hayward
and Bardeen BH metrics. In both cases, we have obtained a family of rotating BH 
spacetimes. For the choice of the complexification in the Newman-Janis prescription, 
we have only required to recover the Kerr solution in the limit of vanishing $g$. 
The remaining freedom is not an ambiguity of the procedure, but it reflects the fact 
there is an infinite number of possible configurations of the matter field (if we see
these metrics as solutions of Einstein's equations with an exotic matter field rather
than solutions of modified Einstein's equations). Our results can also be extended 
to other BH solutions. Generally speaking, there exists a privileged (and trivial, as 
it alters the $1/r$ term only) complexification (type-I solution): in this special case, 
the rotating solution looks like the Kerr metric in Boyer-Lindquist coordinates,
with the mass $M$ promoted to a function $m(r)$ of the $r$ coordinate only.
Independently of the exact form of $m$, the spacetime is of Petrov type~D and the
motion of a free particle is characterized by the existence of the Carter constant. 
For all the other complexifications (type-II solutions), the new metric can be 
written in the Kerr form in null coordinates, with $M$ replaced by some function
 $\tilde{m}(r,\theta)$, whose exact expression depends on the choice of the 
 complexification: here there is no global transformation to recover a solution 
 with a unique non-vanishing off-diagonal coefficient. All the rotating solutions
 generated by the Newman-Janis algorithm from the Hayward and Bardeen 
 metrics seem to be singularity free. However,
 the weak energy condition, satisfied for $a=0$, is now violated. These
 solutions have also a questionable behavior of the curvature invariant at the 
 origin. We have thus outlined the possibility that the Newman-Janis algorithm
 should be generalized by introducing a spin parameter function of the radial
 coordinate. Such a possibility sounds quite natural, as all the regular BH
 solutions have a mass function that may be seen as the mass inside the sphere
 of radius $r$. Such a modification of the Newman-Janis algorithm avoids
 the suspicious behavior of the curvature invariants at the origin and
 makes the rotating solution violate the weak energy condition only by a 
 small amount. We cannot really exclude that with a proper choice of $a'$ in
 Eq.~(\ref{eq-ap}) the violation of the weak energy condition can be completely
 avoided, but all our attempts to do it failed.


\begin{acknowledgments}
We thank David Garfinkle, Hideo Kodama, Luca Lusanna, Kei-ichi Maeda, 
Daniele Malafarina, and Euro Spallucci for useful discussions and suggestions.
This work was supported by the Thousand Young Talents Program 
and Fudan University.
\end{acknowledgments}



\begin{thebibliography}{99}

\bibitem{sing} 
  S.~W.~Hawking and R.~Penrose,
  Proc.\ Roy.\ Soc.\ Lond.\ A {\bf 314}, 529 (1970);
  S.~W.~Hawking and G.~F.~R.~Ellis,
  {\it The Large scale structure of space-time}
  (Cambridge University Press, Cambridge, UK, 1973).
  
\bibitem{mathur} 
  S.~D.~Mathur,
  Fortsch.\ Phys.\  {\bf 53}, 793 (2005)
  [hep-th/0502050].

\bibitem{gia} 
  G.~Dvali and C.~Gomez,
  arXiv:1112.3359 [hep-th];
  arXiv:1203.6575 [hep-th];
  arXiv:1212.0765 [hep-th].

\bibitem{bar} 
  J.~M.~Bardeen,
  in {\it Conference Proceedings of GR5} (Tbilisi, USSR, 1968), p. 174.  
  
\bibitem{ag} 
  E.~Ayon-Beato and A.~Garcia,
  Phys.\ Lett.\ B {\bf 493}, 149 (2000)
  [gr-qc/0009077].
  
\bibitem{hay} 
  S.~A.~Hayward,
  Phys.\ Rev.\ Lett.\  {\bf 96}, 031103 (2006)
  [gr-qc/0506126].

\bibitem{rbh} 
  I.~Dymnikova,
  Gen.\ Rel.\ Grav.\  {\bf 24}, 235 (1992);
  Class.\ Quant.\ Grav.\  {\bf 21}, 4417 (2004)
  [gr-qc/0407072];
  E.~Ayon-Beato and A.~Garcia,
  Phys.\ Rev.\ Lett.\  {\bf 80}, 5056 (1998)
  [gr-qc/9911046];
  K.~A.~Bronnikov,
  Phys.\ Rev.\ D {\bf 63}, 044005 (2001)
  [gr-qc/0006014];
  K.~A.~Bronnikov and J.~C.~Fabris,
  Phys.\ Rev.\ Lett.\  {\bf 96}, 251101 (2006)
  [gr-qc/0511109];
  W.~Berej, J.~Matyjasek, D.~Tryniecki and M.~Woronowicz,
  Gen.\ Rel.\ Grav.\  {\bf 38}, 885 (2006)
  [hep-th/0606185].

\bibitem{rev} 
  C.~Bambi,
  Mod.\ Phys.\ Lett.\ A {\bf 26}, 2453 (2011)
  [arXiv:1109.4256 [gr-qc]];
  Astron.\ Rev.\  {\bf 8}, 4 (2013)
  [arXiv:1301.0361 [gr-qc]].

\bibitem{bb} 
  C.~Bambi and E.~Barausse,
  Astrophys.\ J.\  {\bf 731}, 121 (2011)
  [arXiv:1012.2007 [gr-qc]].

\bibitem{b} 
  C.~Bambi,
  Phys.\ Rev.\ D {\bf 83}, 103003 (2011)
  [arXiv:1102.0616 [gr-qc]];  
  Mod.\ Phys.\ Lett.\ A {\bf 26}, 2453 (2011)
  [arXiv:1109.4256 [gr-qc]];
  Phys.\ Lett.\ B {\bf 705}, 5 (2011)
  [arXiv:1110.0687 [gr-qc]];
  Phys.\ Rev.\ D {\bf 85}, 043002 (2012)
  [arXiv:1201.1638 [gr-qc]];
  Phys.\ Rev.\ D {\bf 86}, 123013 (2012)
  [arXiv:1204.6395 [gr-qc]];
  JCAP {\bf 1209}, 014 (2012)
  [arXiv:1205.6348 [gr-qc]];
  Astrophys.\ J.\  {\bf 761}, 174 (2012)
  [arXiv:1210.5679 [gr-qc]];
  Phys.\ Rev.\ D {\bf 87}, 023007 (2013)
  [arXiv:1211.2513 [gr-qc]].
  
\bibitem{slow} 
  P.~Pani and V.~Cardoso,
  Phys.\ Rev.\ D {\bf 79}, 084031 (2009)
  [arXiv:0902.1569 [gr-qc]];
  N.~Yunes and F.~Pretorius,
  Phys.\ Rev.\ D {\bf 79}, 084043 (2009)
  [arXiv:0902.4669 [gr-qc]];
  K.~Yagi, N.~Yunes and T.~Tanaka,
  Phys.\ Rev.\ D {\bf 86}, 044037 (2012)
  [arXiv:1206.6130 [gr-qc]].
  
\bibitem{degb} 
  B.~Kleihaus, J.~Kunz and E.~Radu,
  Phys.\ Rev.\ Lett.\  {\bf 106}, 151104 (2011)
  [arXiv:1101.2868 [gr-qc]].
  
\bibitem{loop}  
  A.~Smailagic and E.~Spallucci,
  Phys.\ Lett.\ B {\bf 688}, 82 (2010)
  [arXiv:1003.3918 [hep-th]];
  F.~Caravelli and L.~Modesto,
  Class.\ Quant.\ Grav.\  {\bf 27}, 245022 (2010)
  [arXiv:1006.0232 [gr-qc]].
  
\bibitem{belt} 
  L.~Modesto and P.~Nicolini,
  Phys.\ Rev.\ D {\bf 82}, 104035 (2010)
  [arXiv:1005.5605 [gr-qc]].

\bibitem{nj} 
  E.~T.~Newman and A.~I.~Janis,
  J.\ Math.\ Phys.\  {\bf 6}, 915 (1965);
  E T.~Newman, R.~Couch, K.~Chinnapared, A.~Exton, A.~Prakash and R.~Torrence,
  J.\ Math.\ Phys.\  {\bf 6}, 918 (1965).

\bibitem{ds} 
  S.~P.~Drake and P.~Szekeres,
  Gen.\ Rel.\ Grav.\  {\bf 32}, 445 (2000)
  [gr-qc/9807001].
  
\bibitem{teu} 
  J.~M.~Bardeen, W.~H.~Press and S.~A.~Teukolsky,
  Astrophys.\ J.\  {\bf 178}, 347 (1972).
  
\bibitem{wald} 
  R.~M.~Wald,
  {\it General Relativity}
  (Chicago University Press, Chicago, US, 1984).

\end{thebibliography}
\end{document}